\documentclass[prl,twocolumn,showpacs,amsmath,superscriptaddress,psfig,amssymb]{revtex4}

\usepackage{graphicx}
\usepackage{dcolumn}
\usepackage{bm}

\begin{document}


\title{Multiband Superconductivity of Heavy Electrons in TlNi$_2$Se$_2$ single crystal}

\author{Hangdong Wang}
\affiliation {Department of Physics, Zhejiang University, Hangzhou 310027, China}
\affiliation {Department of Physics, Hangzhou Normal University, Hangzhou 310036, China}

\author{Chiheng Dong}
\affiliation {Department of Physics, Zhejiang University, Hangzhou 310027, China}

\author{Qianhui Mao}
\affiliation {Department of Physics, Zhejiang University, Hangzhou 310027, China}

\author{Rajwali Khan}
\affiliation {Department of Physics, Zhejiang University, Hangzhou 310027, China}

\author{Xi Zhou}
\affiliation {Department of Physics, Zhejiang University, Hangzhou 310027, China}

\author{Chenxia Li}
\affiliation {Department of Physics, Zhejiang University, Hangzhou 310027, China}

\author{Bin Chen}
\affiliation {Department of Physics, Hangzhou Normal University, Hangzhou 310036, China}

\author{Jinhu Yang}
\affiliation {Department of Physics, Hangzhou Normal University, Hangzhou 310036, China}

\author{Qiping Su}
\affiliation {Department of Physics, Hangzhou Normal University, Hangzhou 310036, China}

\author{Minghu Fang}
\email{mhfang@zju.edu.cn}
\affiliation {Department of Physics, Zhejiang University, Hangzhou 310027, China}

\date{\today}

\begin{abstract}
\noindent Superconductivity has been first observed in TlNi$_2$Se$_2$ at \textit{T}$_C$=3.7 K and appears to involve heavy electrons with an effective mass $m^*$=14$\sim$20 $m_b$, as inferred from the normal state electronic specific heat and the upper critical field, $H_{C2}(T)$. Although the zero-field electronic specific heat data, $C_{es}(T)$, in low temperatures ($T < 1/4 T_C$) can be fitted with a gap BCS model, indicating that TlNi$_2$Se$_2$ is a fully gapped superconductor, the two-gap BCS model presents the best fit to all the $C_{es}(T)$ data below $T_C$. It is also found that the electronic specific heat coefficient in the mixed state, $\gamma_N(H)$, exhibits a \textit{H}$^{1/2}$ behavior, which was also observed in some \textit{s}-wave superconductors, although once considered as a common feature of the \textit{d}-wave superconductors. Anyway, these results indicate that TlNi$_2$Se$_2$, as a non-magnetic analogue of TlFe$_x$Se$_2$ superconductor, is a multiband superconductor of heavy electron system.

\end{abstract}

\maketitle

The standard heavy fermion compounds containing Ce, Yb and U ions undergo a continuous transition from a high temperature phase in which the \textit{f}-electrons behave as if they are localized to a low temperature heavy Feimion liquid phase in which the \textit{f}-electrons appear to be delocalized with enormous effective masses $m^*$\cite{Fisk 1986,Heffner 1996,Stewart 1984}. The heavy Fermi liquid ground state is unstable to the formation of superconductivity and magnetically ordered states. The superconductivity in these materials appears to be anisotropic with an energy gap with point or line nodes, indicative of superconducting electron pairing with angular momentum greater than zero (\textit{i.e.} \textit{d}-wave)\cite{Fisk 1986,Heffner 1996,Stewart 1984}. It is widely believed that the pairing of the superconducting electrons is mediated by magnetic fluctuations.

An intriguing possibility is the existence of charge order, rather than the usual magnetic order, in proximity to the heavy-fermion state. The materials KNi$_2$Se$_2$ \cite{Neilson 2012} and KNi$_2$S$_2$ \cite{Neilson 2012-1}, in which Ni ion has a mixed valance Ni$^{+1.5}$, have recently been shown to exhibit several remarkable physical properties. At high temperatures they have high resistivity; the magnetic susceptibility is constant; and structural analysis reveals that they have at least three distinct sub-populations of Ni-Ni bond lengths. Upon cooling below $T_{coh}$ $\sim$ 20K, the resistivity rapidly decreases, and the system enters a coherent heavy-fermion state with effective electron mass $m^*$ $\sim$ 10$m_b$, eventually giving way to superconductivity below $T_c$ $\sim$ 1 K. It was suggested \cite{Neilson 2012,Murray 2012} that the formation of a heavy-fermion state at low temperatures is driven by the hybridization of localized charges with conduction electrons, the coherent state competes with a \textit{charge}-fluctuating state, facilitated by the mixed valency of Ni ions in this system. It raises a question whether the superconductivity in this system is unconventional (\textit{i.e.}, \textit{d}-wave), as that in the standard heavy fermion compounds, or conventional, as that in the Ni-pnictide compounds, such as LaNiAsO (\textit{T$_C$}=2.75K) \cite{Li 2008}, BaNi$_2$As$_2$ (\textit{T$_C$}=0.7 K) \cite{Ronning 2008} or SrNi$_2$P$_2$ (\textit{T$_C$}=1.4 K) \cite{Ronning 2009}. Due to an un-stability in air, a relative lower \textit{T$_C$}, and the Schottky anomaly corresponding to impurity in the specific heat for this new Ni-chalcogenide superconductors, KNi$_2$Se$_2$ ($T_C$=0.8 K) and KNi$_2$S$_2$ ($T_C$=0.46 K) polycrystalline samples, there have been few reports on the nature of superconductivity.

TlNi$_2$Se$_2$ crystalizes in a tetragonal ThCr$_2$Si$_2$-type structure (space group $I4/mmm$), shown in Fig. 1(a), the same as that of KNi$_2$Se$_2$, Fe-arsenide superconductors, such as (Ba,K)Fe$_2$As$_2$ \cite{Rotter 2008}, BaFe$_{2-x}$Co$_x$As$_2$ \cite{Sefat 2008}, Co-based superconductor LaCo$_2$B$_2$ \cite{Mizoguchi 2011}, as well as the fist heavy-fermion superconductor CeCu$_2$Si$_2$ \cite{Steglich 1979}. It can be considered one of a non-magnetic analogue of Fe-chalcogenide superconductors recent discovered by us, \textit{i.e.} TlFe$_x$Se$_2$ compounds with Fe vacancies \cite{Fang 2011,Wang 2011}. TlNi$_2$Se$_2$ compound is a Pauli paramagnetic metal, reported first by A.R. Newmark \cite{Newmark 1989}, who did not observe any superconducting transition above 2 K. In this letter, we grew successfully TlNi$_2$Se$_2$ single crystal and rechecked its structure and physical properties. Superconductivity has first been observed in TlNi$_2$Se$_2$ at \textit{T}$_C$=3.7 K and appears to involve heavy electrons with an effective mass $m^*$=14$\sim$20 $m_e$. The zero-field electronic specific heat data, $C_{es}(T)$, in low temperatures ($T < 1/4 T_C$) can be described by $C_{es}$ =$C_0$\textit{exp}($-\triangle$)/$k_B$\textit{T} with $\triangle$ = 3.03K, indicating that TlNi$_2$Se$_2$ is a fully gapped superconductor, similar to that in the other Ni-arsenide compounds, but different with that in the standard heavy fermion \textit{f}-electron compounds. Furthermore, the two-gap BCS model presents the best fit to $C_{es}(T)$ data below $T_C$, illustrating that TlNi$_2$Se$_2$ is also a multi-band superconductor. Another, it is surprising to find that the electronic specific heat coefficient in the mixed state, $\gamma_N(H)$, exhibits a \textit{H}$^{1/2}$ behavior, which was once considered as a common feature of the \textit{d}-wave superconductors, although also observed in some conventional \textit{s}-wave superconductors. These results indicate that TlNi$_2$Se$_2$ is an example of multiband superconductor of heavy electron system.

\begin{figure}
  \includegraphics[width=8cm]{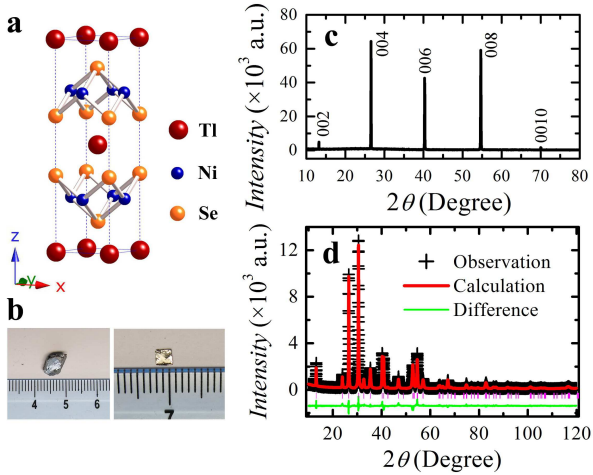}\\
  \caption{(Color online)(a) Crystal structure of stoichiometric TlNi$_2$Se$_2$ with the tetragonal ThCr$_2$Si$_2$-type. (b) Photos of TlNi$_2$Se$_2$ crystals (before being cleaved) and TlNi$_2$Se$_2$ crystal. (c)Single-crystal XRD pattern of TlNi$_2$Se$_2$ and (d) XRD pattern of powder obtained by grinding TlNi$_2$Se$_2$ crystals. Its Rietveld refinement is shown by the red line.}\label{}
\end{figure}

Single crystals of TlNi$_2$Se$_2$ were grown using a self-flux method. A mixture with a ratio of Tl:Ni:Se=1:2:2 was placed in an alumina crucible, sealed in an evacuated quartz tube, heated at 950$^o$C for 12 hours, and cooled to 700$^o$C at a rate of 6$^o$C/h, followed by furnace cooling. In each step to prepare the sample, we managed carefully it due to Tl metal poison. Single crystals with a typical dimension of 2$\times$2$\times$0.2 $mm^3$ [see Fig. 1(b)], were mechanically isolated from the flux. Energy Dispersive X-ray Spectrometer (EDXS) was used to determine the crystal composition, and stoichiometric TlNi$_2$Se$_2$ was confirmed, which is different with the analogue of Fe-chalcogenide compound, such as TlFe$_x$Se$_2$ compound, in which Fe-vacancies existing always \cite{Fang 2011}. The exact composition of TlNi$_2$Se$_2$ indicates that Ni ions should be a mixed valance of Ni$^{1.5+}$ due to Tl having monovalence. X-ray diffraction (XRD) pattern [see Fig. 1(d)] at room temperature of the TlNi$_2$Se$_2$ powder by grinding pieces of crystals confirms its ThCr$_2$Si$_2$-type structure, and its Rietveld refinement (reliability factor: $R_{wp}$= 9.53$\%$, $\chi^2$= 2.671) gives the lattice parameters of \textit{a} = 3.87($\pm$0.01) {\AA} and \textit{c} = 13.43($\pm$0.01) {\AA} and an evidence of no vacancies existing at each sites. Electrical resistivity both \textit{in}-plane ($\rho_{ab}$) and \textit{out-of}-plane ($\rho_c$), specific heat \textit{C}, and magnetic susceptibility $\chi$ measurements between 0.5 K and 300K were made using a Quantum Design MPMS or PPMS.

\begin{figure}
  \includegraphics[width=8cm]{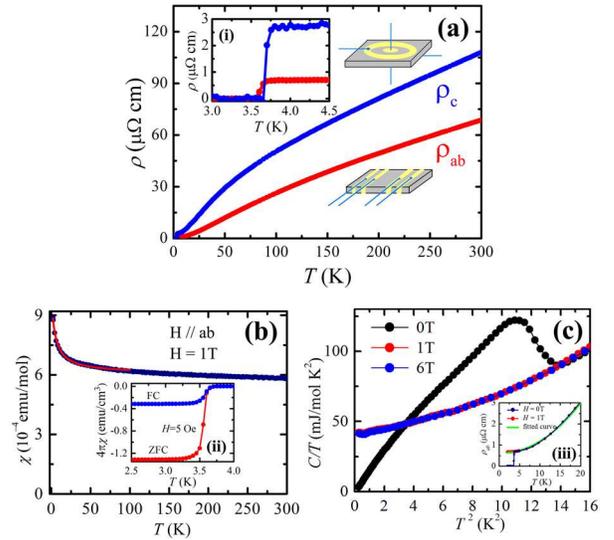}\\
  \caption{(Color online)(a) Temperature dependence of both \textit{in}  and \textit{out-of}-plane resistivity, $\rho$$_{ab}$(T) and $\rho$$_c$(T) for TlNi$_2$Se$_2$ single crystal. (b) Temperature dependence of the normal state magnetic susceptibility, $\chi$(T), measured at 1 Tesla field parallel to \textit{ab} plane. (c) Specific heat divided by temperature, $C/T$ vs $T^2$, measured under 0, 1 and 6 Tesla field. Inset: (i) $\rho$$_{ab}$(T) and $\rho$$_c$(T) near the superconducting transition, (ii) $\chi$(T) near the superconducting transition, measured at 5 Oe field parallel to \textit{ab} plane (for minimizing the demagnetization factor) with both zero-field cooling (ZFC) and field cooling (FC) processes, (iii) $\rho$$_{ab}$(T) below 20 K measured under 0 and 1 Tesla, the green line indicates the fitting line using Fermi-liquid behavior. }\label{}
\end{figure}

The physical properties of TlNi$_2$Se$_2$ are summarized in Fig. 2. Both $\rho_{ab}$ and $\rho_c$ vs \textit{T} curves, shown in Fig. 2(a) and inset (i), display a metallic behavior in the normal state before dropping abruptly to zero when superconductivity occurs at $T_C$=3.7 K, which is also confirmed by a large diamagnetic signal [see the inset (ii) of Fig. 2] and a specific heat jump at $T_C$ as shown in Fig. 2(c). At first, we discuss the resistivity in the normal state. $\rho_{ab}$ and  $\rho_c$ at 300K is of 68.69 and 108.10 $\mu\Omega cm$, respectively, and $\rho_c$/$\rho_{ab}$= 1.57, indicating that the anisotropy in TlNi$_2$Se$_2$ is rather small, although the compound has a layer structure. In order to get $\rho_{ab}$(T) behavior at low temperatures, we measured $\rho_{ab}(T)$ at various magnetic fields above the upper critical field [$\mu_0$$H_{c2}$(0) = 0.802 Tesla, discussed below]. It was found that no change in $\rho_{ab}(T)$ measured at $\mu_0H$$\geq$ 1 up to 6 Tesla occurs, indicating no magneto-resistance response in the normal state. Another, we found that $\rho_{ab}(T)$ data below 25K in the normal state can be very well described by a Fermi-liquid behavior, \textit{i.e.} $\rho_{ab}(T)$=$\rho_0$+$AT^2$. $\rho_0$ =0.615 $\mu\Omega$ cm and \textit{A} =4.94$\times$10$^{-3}$ $\mu\Omega$ cm/K$^2$ were obtained by fitting $\rho_{ab}(T)$ data measured at 1 Tesla [see the red line in the inset (iii) of Fig. 2]. The residual resistivity ratio [RRR =$\rho_{ab}$(300K)/$\rho_{ab}$(2K)$\sim$ 103, where $\rho_{ab}$(2K) obtained from $\rho_{ab}(T)$ data measured at 1 Tesla] and superconducting transition width $\bigtriangleup$$T_C$= 0.05 K reflect the high quality of the single crystals. It is worth to note that no discontinuous change in both $\rho_{ab}(T)$ and $\rho_c(T)$ was observed, which occurs in the iso-structural both BaNi$_2$As$_2$ \cite{Kurita 2009} and SrNi$_2$P$_2$ \cite{Ronning 2009} compounds, corresponding to the structural transition from a tetragonal at higher temperatures to a triclinic at lower temperatures. The powder XRD results (not shown in the paper) at low temperatures provide also an evidence for that no structural transition occurs in TlNi$_2$Se$_2$ compound below 300 K.

\begin{figure}
  \includegraphics[width=8cm]{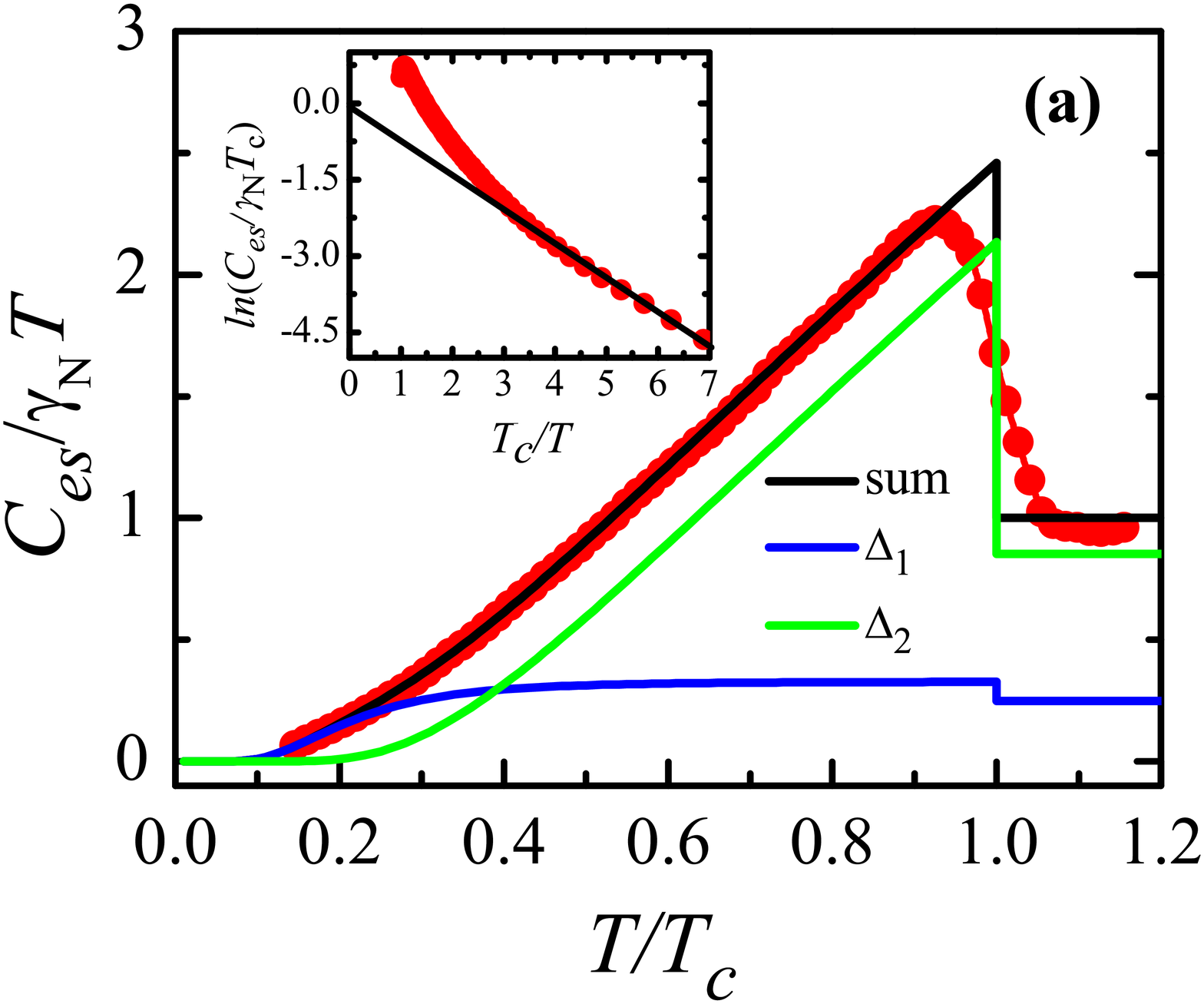}\\
  \includegraphics[width=8cm]{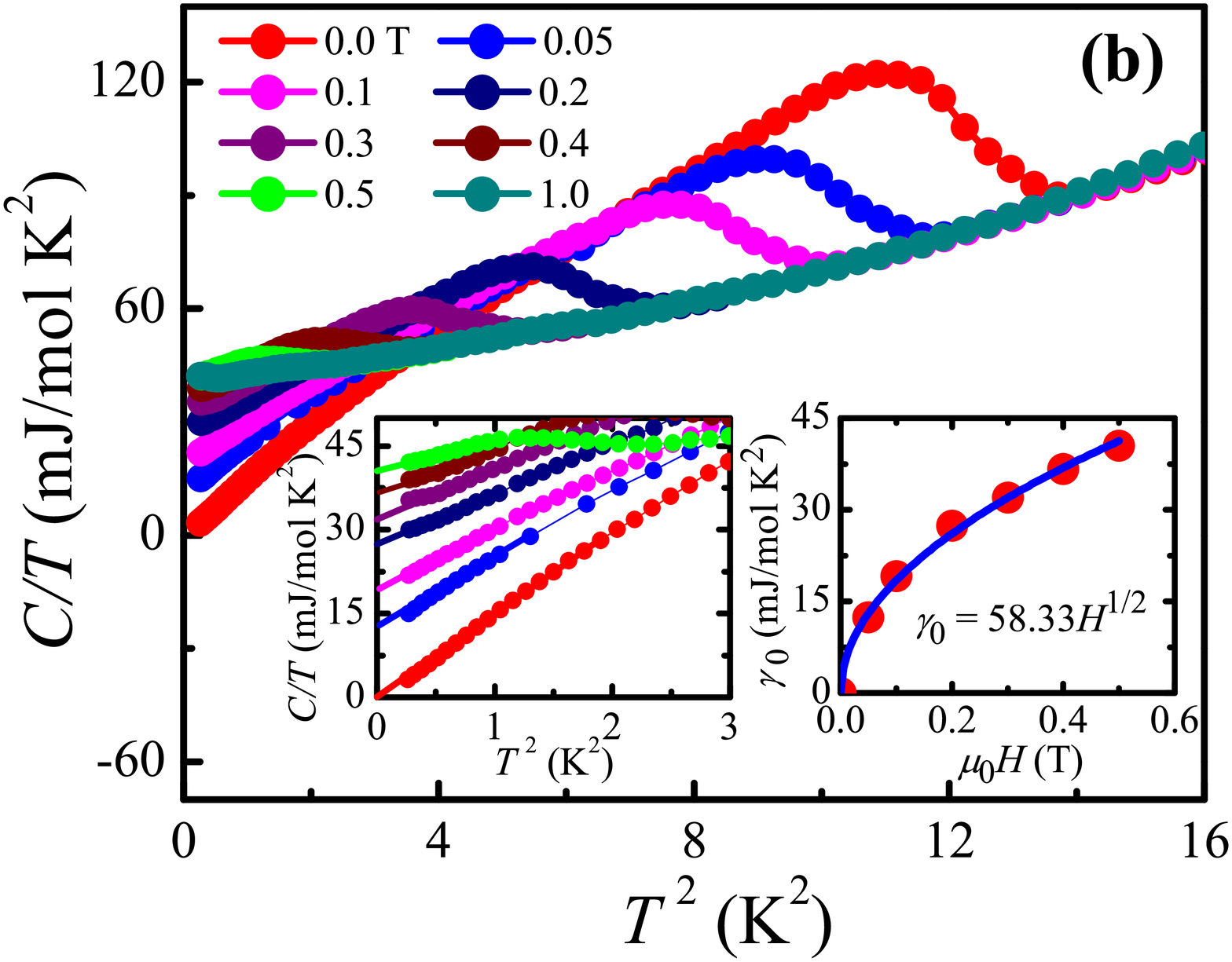}\\
  \caption{(Color online) (a)Reduced temperature, $T/T_C$, dependence of electronic specific heat divided by temperature, $C_{es}$/T, in the superconducting state at zero field, where $C_{es}$=$C-C_{Latt}$. Blue, green and black solid lines show the individual and total contributions of the two gaps to $C_{es}$/\textit{T}, respectively. Inset: Reduced reciprocal temperature, $T_C/T$, vs reduced electronic specific heat, $C_{es}/\gamma_N$$T_C$. (b) Low temperature specific heat divided by temperature, $C/T$, vs $T^2$, measured at various fields near superconducting transition. Left inset: $C/T$, vs $T^2$ blow 1.7K. Right inset: Magnetic field $\mu_0$H dependence of electronic specific heat coefficient $\gamma_0$ in the mixed state .}\label{}
\end{figure}

\begin{figure}
  \includegraphics[width=8cm]{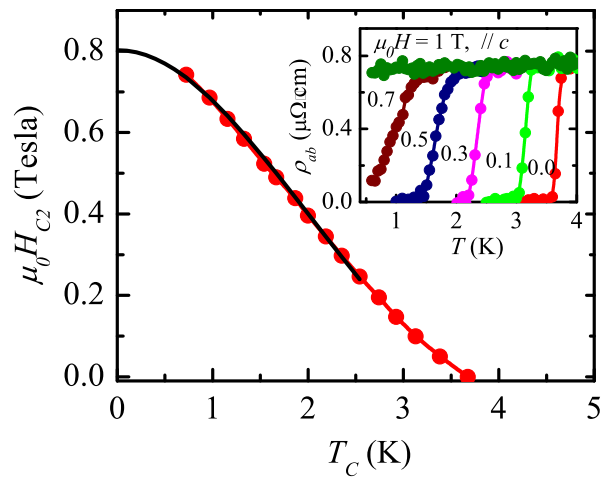}\\
  \caption{(Color online) Upper critical field $H_{C2}$ vs $T$ of TlNi$_2$Se$_2$. Inset: \textit{in}-plane reistivity, $\rho_{ab}(T)$ in magnetic fields up to 1.0 Tesla applied parallel to $c$ axis }\label{}
\end{figure}

Now, we discuss the normal state specific heat $C_N$(T) measured at a magnetic field $\mu_0$$H$ =6 Tesla, as shown in Fig. 2(c). The $C_N$$(T)/T$ vs $T^2$ plot below 4 K showed a pronounced nonlinear behavior, similar to that observed in KNi$_2$S$_2$ \cite{Neilson 2012-1}. We therefore fitted $C_N$(T) with $C_N$(T)= $\gamma_N$$T$+$\beta$$T^3$+$\delta$$T^5$, which yields $\gamma_N$=40 mJ/mol K$^2$, $\beta$= 1.65 mJ/mol K$^4$ and $\delta$= 0.135 mJ/mol K$^6$. The Debye temperature $\Theta_D$ was estimated to be of 175 K. The normal state electronic specific heat coefficient $\gamma_N$ value corresponds to a mass enhancement $m^*/m_b$= 14 (assuming 1.5 carriers/Ni and a spherical Fermi surface). This value is comparable to that of iso-structure KNi$_2$Se$_2$ superconductor (44 mJ/mol K$^2$) \cite{Neilson 2012} and \textit{p}-wave superconductor Sr$_2$RuO$_4$ (40 mJ/mol K$^2$) \cite{Nishizaki 1998}, but is bigger than that in Ni-arsenide superconductors, such as LaO$_{1-x}$F$_x$NiAs (7.3 mJ/mol K$^2$) \cite{Li 2008}, BaNi$_2$As$_2$ (12.3 mJ/mol K$^2$) \cite{Kurita 2009}, while is much smaller than that in the standard heavy-Fermion superconductors, such as CeCu$_2$Si$_2$ (1100 mJ/mol K$^2$) \cite{Steglich 1979}, indicating that the electronic correlation in TlNi$_2$Se$_2$ is stronger than that in Ni-arsenide superconductors, but weaker than that in the standard heavy-Fermion superconductors. The Kadowaki-Woods ratio, $A/\gamma^2$, relates the electronic specific heat to the temperature coefficient in the $\rho_{ab}$(T)=$\rho_0$+\textit{AT}$^2$, and is typically $\sim$10$^{-5}$$\mu\Omega$cm [mol K$^2$mJ]$^2$ for the standard heavy-fermion systems \cite{Kadowaki 1986,Jacko 2009}. For TlNi$_2$Se$_2$, we calculate $A/\gamma^2$ $\sim$ 0.308$\times$10$^{-5}$$\mu\Omega$cm [mol K$^2$mJ]$^2$, which aligns the metallic properties of TlNi$_2$Se$_2$ with heavy-fermion behavior.

As discussed by T. Takayama \textit{et al.} \cite{Takayama 2012} for a new strong-coupling superconductor SrPr$_3$P, we analyze the normal state magnetic susceptibility, $\chi(T)$, shown in Fig. 2(b), measured at a magnetic field of 1 Tesla for TlNi$_2$Se$_2$ crystal. $\chi$=$\chi_0$+$\chi_{CW}$, where $\chi_0$= $\chi_P$+$\chi_{VV}$+$\chi_{core}$ is a temperature independent contribution including a Pauli paramagnetism ($\chi_P$), van Vleck paramagnetism ($\chi_{VV}$) and core diamagnetism ($\chi_{core}$). $\chi_{CW}$ is a Curie-Weiss-like contribution ($<$ 100 K), very likely from magnetic impurities [as shown as the fitting red line in Fig.2(b) corresponding to $<$ 0.97 mol $\%$ of an S=1 impurity, \textit{e.g.} Ni$^{2+}$]. By subtracting the Curie-Weiss-like contribution, we estimate the temperature independent magnetic susceptibility as $\chi_0$= 6.064$\times$10$^{-4}$ emu/mol. The core diamagnetic is estimated to be $\sim$ -0.76$\times$$10^{-4}$ emu/mol by using those reported for Tl$^{1+}$, Ni$^{2+}$ and Se$^{4+}$. Then, the Pauli paramagentic susceptibility is estimated to be $\chi_P$ =6.824$\times$10$^{-4}$ emu/mol by neglecting $\chi_{VV}$, which, combined with $\gamma_N$=40 mJ/mol K$^2$, yields the Wilson ratio $R_W$=$\pi^2$$k_B$$^2$$\chi_P$/3$\mu_B^2$$\gamma_N$ =1.24, typical of many heavy fermion compounds, being comparable to the value of KNi$_2$Se$_2$ ($R_W$=1.71) \cite{Neilson 2012}, larger than than the value ($R_W$=1.0) of the free electron. We note that the Wilson ratio $R_W$ value may be smaller than the value estimated above without subtracting the van Vleck paramagnetism contribution.

Then, we discuss the electronic specific heat in the superconducting state of TlNi$_2$Se$_2$ crystals. Fortunately, no Schottky anomaly in $C(T)$ measurements, at least at $T\geq$0.5 K and $\mu_0H$$\leq$ 6 Tesla, was observed for TlNi$_2$Se$_2$ single crystal, shown in Fig. 2(c), which let us make the analysis on the specific heat data in the superconducting state. Based on the normal state specific heat $C_N$ and $\gamma_N$ obtained above, the electronic specific heat $C_{es}$ in the superconducting state under zero field was estimated as shown in Fig. 3(a), $i.e.$ $C_{es}$=$C(0T)-C_{latt}$, where $C_{latt}$ is phonon contribution by fitting the \textit{C} data measured at 6 T. As fist, we found that low temperature ($T < 1/4 T_C$) specific heat data can be well described by the $C_{es}=C_0exp(-\Delta/k_BT)$, [see the inset of Fig.3(a)], where $k_B$ is the Boltzman constant and the gap $\Delta=3.03 K$ obtained by fitting, indicating that TlNi$_2$Se$_2$ is a fully gapped superconductor, similar to that observed in the other Ni-arsenate compounds \cite{Ronning 2008,Ronning 2009}, but different with that in the standard heavy fermion \textit{f}-electron compounds \cite{Steglich 1979}. Second, it was found that the standard BCS model can not well described all the $C_{es}$ data below $T_C$, while the two-gap BCS model presents the best fit to the $C_{es}$/\textit{T }data [see Fig.3(a)]. According to the phenomenological two-gap model, the heat capacity is taken as the sum of contributions from the two bands, each one following the BCS-type temperature dependence.\cite{Bouquet 2001} In the Fig. 3(a), we plot the contributions from  the two superconducting gaps, $\triangle_1$ = 0.84 $k_BT_C$ and $\triangle_2$ = 2.01 $k_BT_C$, as well as their sum (black line). The weight contributed from the first gap, $\triangle_1$, is about of 0.25. Two gaps behavior in $C_{es}(T)$ is similar to that observed in PrPt$_4$Ge$_{12}$ ($T_C$ $\simeq$ 8K) superconductor. \cite {Zhang 2013}

In order to get much information about its superconducting gap, we also measure the low temperature specific heat at various magnetic fields, $H < 1.0 T$, as shown in Fig. 3(b) and the left inset in Fig. 3(b). At $\mu_0H$ = 0 Tesla, zero linear electronic contribution of $C_{es}$ indicates that almost 100$\%$ electrons enter the superconducting state, different with that (50$\%$) observed in the polycrystalline KNi$_2$Se$_2$ \cite{Neilson 2012} and KNi$_2$S$_2$ \cite{Neilson 2012-1} samples. With increasing magnetic field, the magnitude of specific heat jump at $T_C$ decreases, and the linear electronic specific heat coefficient, $\gamma_N(H)$, obtained by a linear extrapolation of $C_{es}/T$ vs $T^2$ to \textit{T}=0 K, increases. In the mixed state, \textit{i.e.} $H_{C1} < H < H_{C2}$ [where $H_{C1}$ is the lower critical field, $H_{C1}(0)\sim$ 170 Oe for TlNi$_2$Se$_2$, determined by the magnetic hysteresis, $M(H)$, measurements at different \textit{T}$<$ $T_C$, not shown in the paper], the electronic contribution to specific heat is usually attributed to the normal state electrons in the core of vortex. For the \textit{s}-wave superconductors, the cores contribute to $C_{es}$ as a normal metal, then it should be in proportion to the numbers of the cores, \textit{i.e.} $\gamma_N(H)$= $\gamma_0H/H_{C2}$. However, we found that $\gamma_N(H)$=58.33$H^{1/2}$, as shown in the right inset of Fig. 3(b), obtained by fitting the $\gamma_N(H)$ data, which was generally observed in \textit{d}-wave cuprate superconductors \cite{Wright 1999,Yang 2001} and some heavy-fermion compounds, such as UPt$_3$ \cite{Meulen 1990} due to the importance of the Doppler shift to \textit{d}-wave superconductivity. The $\gamma_N(H)$$\propto$ $H^{1/2}$ behavior was once considered as a common feature of the \textit{d}-wave superconductors. In fact, this behavior was also observed in other \textit{s}-wave superconductors, such as NbSe$_2$ \cite{Sanchez 1995,Sonier 1999}, V$_3$Si \cite{Ramirez 1996} and CeRu$_2$ \cite{Hedo 1998} and the organic superconductor (BEDT-TTF)$_2$Cu[N(CN)$_2$]Br \cite{Nakazawa 1997} and the boroncarbide superconductor LuNi$_2$B$_2$C \cite{Nohara 1997}. Ramirez \cite{Ramirez 1996} suggested that this behavior must be a general feature of all superconductors in the vortex state, independent of the order parameter symmetry, but somehow related the strength of the vortex-vortex interactions.

Finally, we present the analysis of the upper critical field $H_{C2}(T)$, which give an evidence for superconductivity emerging from the heavy-mass electronic state in TlNi$_2$Se$_2$, similar to that described for the standard heavy fermion superconductor, such as for UBe$_{13}$ \cite{Maple 1985}. Shown in Fig.4 is the $H_{C2}(T)$ curve for TlNi$_2$Se$_2$ derived from $\rho_{ab}$(T) measurements (inset of Fig.3) as a magnetic field applied parallel to \textit{c} axis. Using the middle superconducting transition temperature, the zero temperature upper critical field $H_{C2}$(0) can be estimated with a formula $H_{C2}(T)= H_{C2}(0)(1-t^2)/(1+t^2)$ \cite{Tinkham 1975,Li 2008}, where \textit{t} is the reduced temperature $t=T/T_C$, yielding the value of $\mu_0H_{C2}$(0)= 0.802 Tesla by fitting [see the main panel of Fig. 4]. The superconducting coherence length $\xi_0$ can be estimated from the relation $\xi_0$=[$\Phi_0$/2$\pi$$H_{c2}$]$^{1/2}$, yielding  $\xi_0$$^{ab}$=20.3 nm. A value for the Fermi velocity $v_F$=5.484$\times$10$^4$ m/s is then obtained from  $\xi_0$=0.18$\hbar$$v_F$/$k_B$$T_C$ \cite{Tinkham 1975} from which $m^*$ and $\gamma_N$ can be estimated as follows. Using a spherical Fermi surface approximation, the Fermi wave vector is given by $k_F$=(3$\pi^2$Z/$\Omega$)$^{1/3}$, where \textit{Z} is the number of electrons per unit cell and $\Omega$ is the unit cell volume. Assuming that Ni contributes 1.5 electrons (\textit{Z}=6), we obtain $k_F$=9.6$\times$10$^9$ m$^{-1}$. The expression $m^*$=$\hbar$$k_F$/$v_F$ yields m*$\sim$ 20 m$_e$. From the relation $\gamma_N$=$\pi^2$$N$$k_B^2$$m^*$/$\hbar^2$$k_F^2$, $\gamma_N$ $\sim$ 61 mJ/mol K$^2$. The values of $m^*$ and $\gamma_N$ are comparable to the values estimated from the normal state specific heat.

In summary, TlNi$_2$Se$_2$ exhibits superconductivity with $T_C$=3.7 K that appears to involve heavy electrons with an effective mass $m^*$=14$\sim$20 $m_b$, as inferred from the electronic specific heat coefficient $\gamma_N$ and the upper critical field, $H_{C2}(0)$. The zero-field electronic specific heat data, $C_{es}(T)$, in low temperatures ($T < 1/4 T_C$) can be fitted by a one gap BCS model, indicating that TlNi$_2$Se$_2$ is a fully gapped superconductor, similar to that in the other Ni-arsenide compounds, but different with that in the standard heavy fermion \textit{f}-electron compounds. But the two-gap BCS model presents the best fit to $C_{es}(T)$ data below $T_C$. It is also found that the electronic specific heat coefficient in the mixed state, $\gamma_N(H)$, exhibits a \textit{H}$^{1/2}$ behavior, which was once considered as a common feature of the \textit{d}-wave superconductors, but also observed in some conventional \textit{s}-wave superconductors. Anyway, these results indicate that TlNi$_2$Se$_2$, as a non-magnetic analogue of TlFe$_x$Se$_2$ superconductor, is a multiband superconductor of heavy electron system.

This work is supported by the National Basic Research Program of China (973 Program) under grant No. 2011CBA00103, 2012CB821404 and 2009CB929104, the Nature Science Foundation of China (Grant No. 10974175, 10934005 and 11204059) and Zhejiang Provincial Natural Science Foundation of China (Grant No. Q12A040038), and the Fundamental Research Funds for the Central Universities of China. We thank Dr. Xiaofeng Xu for helpful discussions.

\end{document}